 \documentclass[manuscript]{emulateapj}

 \usepackage{apjfonts}

\slugcomment{to appear in the Astrophysical Journal, Letters}

\shorttitle{Light Curve Models of Novae}
\shortauthors{Hachisu \& Kato}

\begin{document}

\title{Optical and Supersoft X-ray Light Curve Models of
Classical Nova V2491 Cygni: A New Clue to the Secondary Maximum}


\author{Izumi Hachisu}
\affil{Department of Earth Science and Astronomy, 
College of Arts and Sciences, University of Tokyo,
Komaba, Meguro-ku, Tokyo 153-8902, Japan} 
\email{hachisu@ea.c.u-tokyo.ac.jp}

\and

\author{Mariko Kato}
\affil{Department of Astronomy, Keio University, 
Hiyoshi, Kouhoku-ku, Yokohama 223-8521, Japan} 
\email{mariko@educ.cc.keio.ac.jp}

%




\begin{abstract}
     V2491 Cygni (Nova Cygni 2008 No.2) was detected as a transient
supersoft X-ray source with the {\it Swift} XRT as early as 40 days
after the outburst, suggesting a very massive white dwarf (WD) close
to the Chandrasekhar limit.  We present a unified model
of near infrared, optical, and X-ray light curves for V2491 Cyg,
and have estimated, from our best-fit model,
the WD mass to be $1.3 \pm 0.02~M_\sun$ with
an assumed chemical composition of the envelope,
$X=0.20$, $Y= 0.48$, $X_{\rm CNO} =0.20$,
$X_{\rm Ne} =0.10$, and $Z = 0.02$ by mass weight.
We strongly recommend detailed composition analysis of the ejecta
because some enrichment of the WD matter suggests that
the WD mass does not increase like in RS Oph, which is a
candidate of Type Ia supernova progenitors.  
V2491 Cyg shows a peculiar secondary maximum in the optical
light curve as well as V1493 Aql and V2362 Cyg.
Introducing magnetic activity as an adding energy source
to nuclear burning, we propose a physical mechanism of
the secondary maxima.
\end{abstract}




\section{Introduction}
Classical novae show a wide variety of timescales and shapes in the
optical light curves \citep[e.g.,][]{pay57}.
Among various shapes of nova light curves, V1493 Aql (Nova Aquilae 1999
No.1) shows an impressive secondary maximum about 50 days after the
outburst \citep[e.g.,][]{bon00, ven04}, although
the physical mechanism of the secondary maximum is not understood yet.
Recent two novae, V2362 Cyg (Nova Cygni 2006) and V2491 Cyg
(Nova Cygni 2008 No.2), also show a similar type of single secondary
maximum, at about 250 and 15 days after the outburst, respectively.
These three novae form a wide variety set of timescales, i.e.,
about 15, 50, and 250 days at the secondary maximum and of
secondary peak heights, i.e., 1.1, 2.8, and 3.6 mag, respectively,
\citep[see, e.g.,][]{kim08}, which provide us a new clue
to the mechanism of the secondary
maxima.


In this Letter, we propose a strong magnetic activity as the 
mechanism of the secondary maxima observed in V2491 Cyg, V1493 Aql,
and V2362 Cyg, using the white dwarf (WD) parameters obtained from
light curve fittings based on an optically thick wind model of nova
outbursts \citep{kat94h}.  In \S 2, we briefly describe
our numerical method and light curve fitting of V2491 Cyg.
In \S 3, we propose an idea of strong magnetic activity in the 
WD envelope and estimate the timescales of the secondary maxima.
V1493 Aql and V2362 Cyg show very different
timescales of the secondary maximum, both of which are also explained 
by the same mechanism in \S 4.  Conclusions follow in \S 5.

\section{Modeling of Nova Outbursts}

\subsection{Optically thick wind model}
     After a thermonuclear runaway sets in on a mass-accreting WD,
its photospheric radius expands greatly to $R_{\rm ph} \gtrsim 100 ~R_\sun$
and the WD envelope settles in a steady-state.  We have followed
evolutions of novae by connecting steady state solutions along the
decreasing envelope mass sequence.  We solve a set of equations,
that is, the continuity, equation of motion, radiative diffusion,
and conservation of energy, from the bottom of the hydrogen-rich
envelope through the photosphere assuming spherical symmetry.
Winds are accelerated deep inside the photosphere so that
they are called ``optically thick winds.''
As one of the boundary conditions for our numeral code,
we assume that photons are emitted at the photosphere
as a blackbody with the photospheric temperature of $T_{\rm ph}$.
X-ray flux is estimated directly from the blackbody 
emission, but infrared and optical fluxes are calculated from
free-free emission by using the physical values of our wind solutions.
We neglect the effect of ash helium layer, which may be piled up
beneath the hydrogen burning zone, for all nova calculations except
the RS Oph case \citep{hac07kl}.
Our method and various physical properties of these
wind solutions have already been published \citep[e.g.,][]{hac01ka,
hac01kb, hac04k, hac06kb, hac07k, hkn96, hkn99, hknu99,
hkkm00, hac03a, hac07kl, hac08kc, kat83, kat97, kat99, kat94h}.

     The light curves of our optically thick wind model are
parameterized by the WD mass ($M_{\rm WD}$), 
chemical composition of the envelope ($X$, $Y$, $X_{\rm CNO}$, 
$X_{\rm Ne}$, and $Z$), and the envelope mass 
($\Delta M_{\rm env, 0}$) at the outburst (day 0).
Details of our light curve fittings are described in
\citet{hac06kb, hac07k} and \citet{hac07kl, hac08kc}.


\begin{figure}
\epsscale{1.0}
\plotone{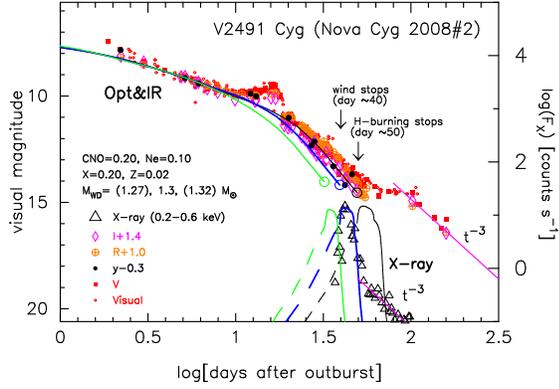}
\caption{
Model light curves of V2491 Cyg for three white dwarf masses,
$1.32~M_\sun$ (green line), $1.3~M_\sun$ 
(thick blue line),  and $1.27~M_\sun$ (thin black line),
together with the observations.  See \citet{hac06kb} for more
details of light curve fitting. Two arrows indicate epochs
when the wind stops (day 40) and when the hydrogen shell-burning
ends (day 50) in the $1.3 ~M_\sun$ WD model.
{\it Dashed lines}: Supersoft X-rays are probably not detected
during the wind phase because of self-absorption by wind itself
\citep[see, e.g.][]{hac03kb}.
Large open circles at the right end of each optical light curve
denote the epoch when the optically thick wind stops.
We obtain the best-fit model for the envelope chemical composition of
$X=0.20$, $Y=0.48$, $X_{\rm CNO} =0.20$, $X_{\rm Ne} =0.10$, and $Z=0.02$.
{\it Large open triangles}: Observational X-ray ($0.2-0.6$ keV)
count rates obtained with {\it Swift} \citep{pag09}.
Optical and near IR observational data of $I$ ({\it open diamonds}),
$R$ ({\it circles with a plus}), $V$ ({\it filled squares}),
$y$ ({\it filled circles}), and visual ({\it small open circles})
are taken from AAVSO (American Association of Variable Star Observers)
and VSOLJ (Variable Star Observers League in Japan).  
The $F_\lambda \propto t^{-3}$ law ({\it magenta}) is added for the
nebular phase, i.e., after the wind stops, where $t$ is the time
after the outburst.
\label{all_mass_v2491_cyg_x20z02o20ne10}}
\end{figure}

\subsection{Light curve fitting (V2491 Cyg)}
     V2491 Cyg was discovered by Nishiyama and Kabashima 
at mag 7.7 on 2008 April 10.728 UT \citep{nak08}.
The nova was not detected on April 8.831 UT (limiting mag 14).
The exact outburst day is unknown, so we assume here that
$t_{\rm OB}= 2454566.0$ (April 9.5 UT) is the outburst day (day 0).
The orbital period of $P_{\rm orb}= 0.0958$~days was derived
by \citet{bak08} from the modulations with an amplitude of $0.03-0.05$ mag.

     The rise or decay time of supersoft X-ray flux is an important
indicator of the WD mass \citep[e.g.,][]{hac06kb, hac07k, hac08kc}.
The best fit model with the {\it Swift} observation \citep{pag09}
is the WD mass of $1.3 \pm 0.02~M_\sun$ as shown
in Figure \ref{all_mass_v2491_cyg_x20z02o20ne10}.  Here we adopt
the chemical composition of $X= 0.20$, $Y=0.48$, $X_{\rm CNO}= 0.20$,
$X_{\rm Ne}= 0.10$, and $Z=0.02$ to reproduce a short (10 days)
supersoft X-ray duration \citep[see][for dependence on
chemical composition]{hac06kb, hac07k, hac07kl, hac08kc}.

Figure \ref{all_mass_v2491_cyg_x20z02o20ne10} also shows
optical and near IR light curves of
the $I$, $R$, $V$, $y$, and visual magnitudes.
Our theoretical light curves of free-free emission reasonably
fit with the observation until day $\sim 50$ except the secondary peak.
After that, the observational data deviate probably due to
the contribution of nebular emission lines.

Thus we conclude that the WD of V2491 Cyg is as massive as 
$1.3 ~M_\sun$ from our light curve fittings with the
supersoft X-ray, optical, and near IR data.
This WD mass of V2491 Cyg is a bit less massive but
comparable to that of the recurrent nova RS Oph \citep{hac07kl},
the supersoft X-ray duration of which lasted 60 days
starting from day 30, much longer than the 10 days in V2491 Cyg.
\citet{hac07kl} explained this long duration of supersoft X-ray
phase of RS Oph by incorporating heat flux from a hot helium 
layer underneath the hydrogen burning zone, which develops
in mass-increasing WDs.  In V2491 Cyg, however, the supersoft
X-ray phase lasts only 10 days, indicating that no thick helium
layer develops beneath the hydrogen burning zone and, therefore,
the WD mass is not increasing \citep[see discussion in][]{hac07kl}.
\citet{tom08} suggested a possibility that V2491 Cyg is
a recurrent nova, for which we expect that the WD mass is increasing
\citep{hac01kb}.
One of the strongest clues to this question is
the chemical composition of the ejecta.  
We strongly recommend composition analysis of V2491 Cyg
to clarify whether or not the ejecta are contaminated by WD matter.
{\it Suzaku} X-ray spectra analyzed by \citet{tak09}
suggested oxygen/neon-rich ejecta.  If so, the WD mass is decreasing and,
as a result, V2491 Cyg is not a candidate of Type Ia supernova
progenitors.


\begin{figure}
\epsscale{1.0}
\plotone{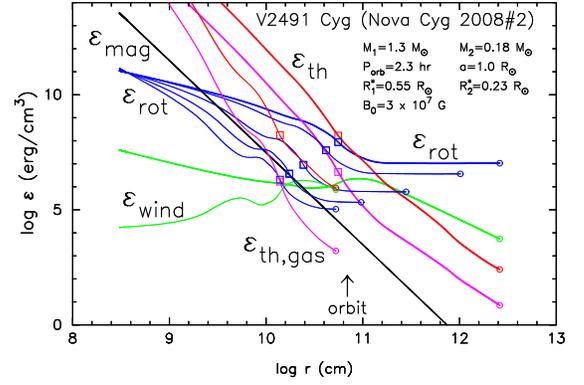}
\caption{
Five energy densities of the WD envelope are plotted against the radius, 
i.e., thermal energy ($\varepsilon_{\rm th}$: {\it red}),
thermal gas energy ($\varepsilon_{\rm th,gas}$: {\it magenta}),
rotational kinetic energy ($\varepsilon_{\rm rot}$: {\it blue}),
magnetic energy ($\varepsilon_{\rm mag}$: {\it black}), and
wind kinetic energy ($\varepsilon_{\rm wind}$: {\it green}).
Five sequential stages are plotted for $\varepsilon_{\rm rot}$ but
only two stages [the first ({\it thick solid}) and the last ({\it thin
solid})] for the other energy densities.
Open circles indicate the photosphere
while open squares correspond to the critical
point of each wind solution.
Various parameters are summarized in the figure.
\label{v2491_cyg_magnetic_field2}}
\end{figure}

\section{Magnetic Activity (V2491 Cyg)}

V2491 Cyg shows a single secondary maximum in the optical and near IR
light curves, which we cannot reproduce by our evolution model.
Here we propose magnetic activities as a new energy source of
the secondary maximum.

The prenova X-ray was detected in V2491 Cyg with {\it Swift}
\citep{iba08a}.  \citet{iba08b} suggested that the prenova 
X-ray spectra are more like those seen from magnetic rather
than non-magnetic cataclysmic variables.  
\citet{tak09} reported the first detection of superhard X-rays
($15-60$ keV) with the {\it Suzaku} HXD at day 10,
the spectrum of which has a power law distribution,
suggesting a non-thermal origin, and unlikely to arise
from thermal emission from internal shocks.
This superhard component was not
seen in their second {\it Suzaku} observation at day 30.
These two observations point toward 
strong magnetic activities on the WD surface.

Therefore, our idea on the secondary maximum is based on
additional energy release associated with rotating magnetic field.
We assume that V2491 Cyg is a polar system with magnetic field
as strong as $B_0 \sim 10^7$ G on the WD surface \citep[e.g.,][]{war95}.
Before the nova outburst, the WD magnetic field rotates synchronously
with the WD spin as well as the binary orbital motion.
After the outburst, the nova envelope expands largely and rotates
differentially due to local angular momentum conservation.  Then,
the magnetic field no more rotates synchronously with the WD spin
because the magnetic tension is not strong enough to keep 
the whole envelope rotating with the WD spin.  We expect that the
differential rotation amplifies magnetic field which drives 
strong magnetic activities.  As the nova outburst proceeds, 
the density of the WD envelope is gradually decreasing
due mainly to wind mass loss and then the magnetic field
eventually recovers synchronous rotation. 
The strong magnetic activities end at this stage, corresponding
to the end of a secondary maximum.  Since the additional energy
source disappears, the light curve
goes back to a ``normal'' one as shown in Figure 
\ref{all_mass_v2491_cyg_x20z02o20ne10}.

     To confirm this idea we estimate
the total thermal energy of gas plus radiation ($\varepsilon_{\rm th}
= \varepsilon_{\rm th,gas} + \varepsilon_{\rm th,rad}$),
rotational kinetic energy $\varepsilon_{\rm rot}$,
wind kinetic energy $\varepsilon_{\rm wind}$, and
magnetic energy $\varepsilon_{\rm mag}$, which are calculated from
\begin{equation}
\varepsilon_{\rm th} = \varepsilon_{\rm th,gas} 
+ \varepsilon_{\rm th,rad} = 
{3 \over 2} {{k T} \over {\mu m_H}} \rho + a T^4,
\end{equation}
\begin{equation}
\varepsilon_{\rm rot} = {1 \over 2} \rho
\left( r~ \Omega_{\rm spin} \right)^2,
\end{equation}
\begin{equation}
\varepsilon_{\rm wind} = {1 \over 2} \rho v_{\rm wind}^2,
\end{equation}
\begin{equation}
\varepsilon_{\rm mag} = {{B_0^2} \over {8 \pi}}
\left( {r \over R_{\rm WD}} \right)^{-4}.
\end{equation}
The temperature $T$, density $\rho$, and wind velocity 
$v_{\rm wind}$ are taken from our
wind solutions of the best-fit model of V2491 Cyg ($1.3 ~M_\sun$ WD).
Here we assume the dipole magnetic
field of $B_0 = 3 \times 10^7$ G at the WD surface
\citep[e.g.,][for V1500 Cyg]{war95}
and that the WD spin period is the same as the
orbital period, $2 \pi / \Omega_{\rm spin}= P_{\rm spin} = P_{\rm orb}$.
Figure \ref{v2491_cyg_magnetic_field2} shows the energy densities
for five sequential stages during the nova outburst.
The first model ({\it thick solid}) has the largest photospheric
radius and corresponds to a stage at/near the optical maximum.
We see that at/near the optical maximum, $\varepsilon_{\rm mag}
\lesssim \varepsilon_{\rm rot}$ at $r \gtrsim 2 \times 10^9$~cm.
This indicates that the magnetic field is differentially
rotating outside of $r \sim 2 \times 10^9$~cm.
Then, the rotation energy density decreases with time and eventually
$\varepsilon_{\rm mag}$ becomes comparable to $\varepsilon_{\rm rot}$
at the stage of the smallest photospheric radius.  After that,
the magnetic field probably gains synchronous rotation with the WD spin. 
We expect that the magnetic activities have a peak at
$\varepsilon_{\rm mag} \approx \varepsilon_{\rm rot}$.
This condition is satisfied at day 15 for our best-fit model,
being very consistent with the time of the secondary maximum.


\begin{deluxetable}{lllllll}
\tabletypesize{\scriptsize}
\tablecaption{Epoch of Secondary maximum in our nova models
\label{secondary_maximum}}
\tablewidth{0pt}
\tablehead{
\colhead{object} &
\colhead{max\tablenotemark{a}} &
\colhead{$P_{\rm orb}$} &
\colhead{$M_2$\tablenotemark{b}} &
\colhead{$a$} &
\colhead{$M_{\rm WD}$} &
\colhead{$\varepsilon_{\rm rot} \approx \varepsilon_{\rm mag}$}\\
\colhead{} &
\colhead{(day)} &
\colhead{(day)} &
\colhead{$(M_\sun$)} &
\colhead{$(R_\sun$)} &
\colhead{$(M_\sun$)} &
\colhead{(day)} 
}
\startdata
V2491 Cyg & 15 & 0.0958 & 0.18 & 1.0 & 1.32 & 13 \\
 &  & & & &  1.3 &  15 \\
 &  & & & &  1.27 & 18 \\
V1493 Aql & 50 & 0.156 & 0.34 & 1.4 & 1.2 & 40 \\
 &  & & & &  1.15 & 49 \\
 &  & & & &  1.1 & 62 \\
V2362 Cyg & 250 & 0.207 & 0.48 & 1.6 & 0.75 & 200 \\
 &  & & & &  0.7 & 240 \\
 &  & & & &  0.65 & 330 
\enddata
\tablenotetext{a}{epoch of secondary maximum}
\tablenotetext{b}{calculated from equation (\ref{warner_mass_formula})}
\end{deluxetable}

     Next we estimate the epoch when the companion emerges from
the nova envelope.   If the mass of the donor star is estimated
from Warner's (1995) empirical formula, i.e.,
\begin{equation}
{{M_2} \over {M_\sun}} \approx 0.065 \left({{P_{\rm orb}}
\over {\rm hours}}\right)^{5/4},
\mbox{~for~} 1.3 < {{P_{\rm orb}} \over {\rm hours}} < 9
\label{warner_mass_formula}
\end{equation}
we have $M_2 = 0.18 ~M_\sun$, which corresponds to
the separation of $a = 1.0 ~R_\sun$,
and the effective Roche lobe radius of the primary component
(WD) of $R_1^* =  0.55 ~R_\sun$.
When the photospheric radius of the nova envelope shrinks
to near the orbit (the separation $a$), 
the condition of $\varepsilon_{\rm rot} < \varepsilon_{\rm mag}$
is satisfied as shown in Figure \ref{v2491_cyg_magnetic_field2}.
This indicates that the epoch of the secondary maximum 
is shortly after the companion emerges from the nova envelope.
We here implicitly assume that strong magnetic field connects the WD and
the companion (like in polar systems).  The mechanism of activity
we suppose is magnetic reconnection.  Strong magnetic reconnection
occurs between the WD and the companion.  When the WD photosphere is
larger than the companion's orbit, magnetic reconnection may occur deep
inside the photosphere but the gas pressure (or gas thermal energy)
at the reconnection region is much larger than the magnetic pressure
(magnetic energy), so the gas is not easily accelerated by magnetic
force.  On the other hand, when the gas pressure becomes smaller than
the magnetic pressure, i.e., when the photosphere shrinks to the orbit
(see Fig.\ref{v2491_cyg_magnetic_field2}), then the gas is easily
accelerated by magnetic force and the envelope gas
is massively ejected.  Thus this process increases the mass-loss rate
around/near the secondary maximum.


\begin{figure}
\epsscale{1.0}
\plotone{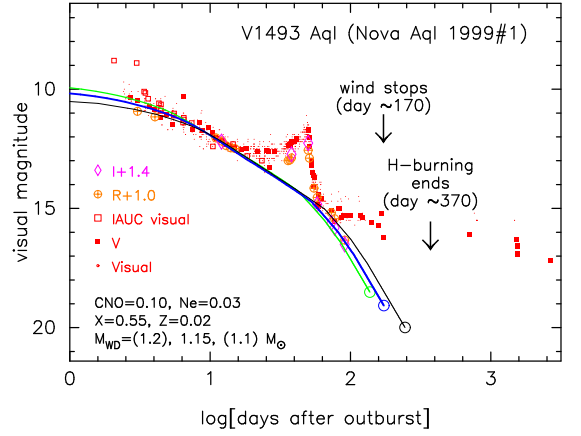}
\caption{
Same as Fig. \ref{all_mass_v2491_cyg_x20z02o20ne10}, but for
V1493 Aql (Nova Aql 1999 No.1).
Our best-fit model is $M_{\rm WD}= 1.15~M_\sun$
for the envelope chemical composition of $X=0.55$, $Y=0.30$,
$X_{\rm CNO} =0.10$, $X_{\rm Ne}=0.03$, and $Z=0.02$.  Observational
data are taken from AAVSO and VSOLJ.  We further add the data of
IAU Circ. 7223, 7225, 7228, 7232, 7258, 7273, 7313.
\label{all_mass_v1493_aql_x55z02o10ne03}}
\end{figure}


\begin{figure}
\epsscale{1.0}
\plotone{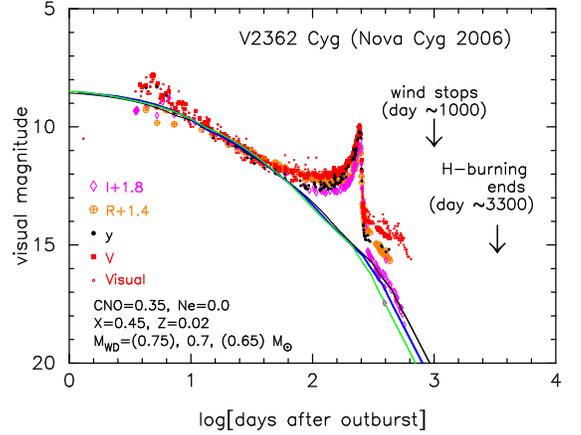}
\caption{
Same as Fig. \ref{all_mass_v2491_cyg_x20z02o20ne10}, but for
V2362 Cyg (Nova Cyg 2006).
Our best-fit model is $M_{\rm WD}= 0.7~M_\sun$
for the envelope chemical composition of $X=0.45$, $Y=0.18$,
$X_{\rm CNO} =0.35$, and $Z=0.02$.  Observational data are
taken from AAVSO and VSOLJ.
\label{all_mass_v2362_cyg_x45z02c15o20}}
\end{figure}

\section{V1493 Aql and V2362 Cyg}

V1493 Aql was discovered by Tago at mag 8.8 on July 13.558 UT.
The nova was not detected on his films of July 5 and 9
(limiting mag 11 and 10.5, respectively).  Therefore, we assume
here that the outburst day is $t_{\rm OB}= 2451372.0$ 
(July 12.5 UT).  Figure \ref{all_mass_v1493_aql_x55z02o10ne03}
shows optical and near infrared light curves of V1493 Aql.
Our theoretical light curves of free-free emission reasonably
fit with the observation until day $\sim 100$ except 
the secondary maximum.  The best-fit model is the WD mass of
$1.15 \pm 0.05~M_\sun$ for the chemical composition of $X=0.55$,
$Y= 0.30$, $X_{\rm CNO}= 0.10$, $X_{\rm Ne}= 0.03$, and $Z=0.02$.
The $V$ and visual magnitudes deviate from our model light curve
about 100 days after the outburst.  This is due probably
to the contribution of emission lines such as [\ion{O}{3}]
in the nebular phase.
The orbital period of $P_{\rm orb}= 0.156$ days was obtained
by \citet{dob06} from the modulations
with a very small amplitude of 0.015 mag.
The mass of the donor star is estimated to be $M_2 = 0.34 ~M_\sun$
from equation (\ref{warner_mass_formula}).  The corresponding 
separation is $a = 1.4 ~R_\sun$ and the effective radius of the WD
Roche lobe is $R_1^* =  0.68 ~R_\sun$.
We have obtained the epoch of $\varepsilon_{\rm mag} \approx
\varepsilon_{\rm rot}$ to be day 49 for 
our $1.15~M_\sun$ WD model as listed in 
Table \ref{secondary_maximum}.
This timescale is consistent with the peak of the secondary
maximum and also the emergence
of the companion star from the WD envelope.

V2362 Cyg was discovered by Nishimura at mag 10.5
on 2006 April 2.807 UT.  The nova was not detected 
on his films taken on March 28 (limiting mag 12) or on earlier patrol
films back to 2001.  Therefore, we assume
here the outburst day of $t_{\rm OB}= 2453827.0$ (April 1.5 UT).
Optical and near infrared light curves are plotted in
Figure \ref{all_mass_v2362_cyg_x45z02c15o20}.
Our theoretical light curves of free-free emission again reasonably
fit with the $I$ observation until day $\sim 500$ except 
the secondary maximum.  The best-fit model is the WD mass of
$0.7 \pm 0.05~M_\sun$ for the chemical composition of
$X=0.45$, $Y= 0.18$, $X_{\rm CNO}= 0.35$, and $Z=0.02$
\citep[see, e.g.,][for observed values]{mun08}.
The $R$ and $y$ magnitudes slightly but 
the $V$ and visual magnitudes largely deviate from our model light curve
after the secondary maximum.  
The orbital period of $P_{\rm orb}= 0.207$ days was obtained
by \citet{gor08} from the modulations with an amplitude of 0.11 mag.
If we take the donor mass of $M_2 = 0.48 ~M_\sun$
from equation (\ref{warner_mass_formula}), 
the separation is $a = 1.6 ~R_\sun$,
and the WD Roche lobe is $R_1^* =  0.66 ~R_\sun$.
We have examined the epoch of $\varepsilon_{\rm mag} \approx
\varepsilon_{\rm rot}$ to be day 240 for our model of $0.7~M_\sun$ WD
as listed in Table \ref{secondary_maximum}.
This timescale is again reasonably consistent with
the peak of the secondary maximum and also the emergence
of the companion star from the WD envelope.

\section{Conclusions}
\label{conclusions}

     We have estimated the WD mass of
the classical novae V2491 Cyg by comparing our free-free
light curves with the optical and near infrared observations
as well as by comparing our blackbody X-ray light curves with
the {\it Swift} XRT data.
The best-fit model is the $1.3 \pm 0.02 ~M_\sun$ WD for the
chemical composition of $X=0.20$, $Y=0.48$,
$X_{\rm CNO}= 0.20$, $X_{\rm Ne}= 0.10$, and $Z=0.02$.
We strongly recommend composition analysis of the ejecta because
the enrichment of WD matter provides information whether the WD
mass increases or not.  We have also estimated the
WD mass of V1493 Aql and V2362 Cyg
to be $M_{\rm WD}= 1.15 \pm 0.05$
and $0.7 \pm 0.05 ~M_\sun$, respectively.
For these three novae, the epoch of secondary maximum
is consistently explained by our magnetic activity model
if the magnetic activity reaches maximum at $\varepsilon_{\rm mag}
\approx \varepsilon_{\rm rot}$ in the WD envelope.
We strongly recommend search for magnetic activities for
these three novae even in quiescence.



\acknowledgments
     We thank D. Takei for providing us with their 
machine readable X-ray data of V2491 Cyg and also 
AAVSO and VSOLJ for the optical and near infrared data for
V2491 Cyg, V1493 Aql, and V2362 Cyg.
We are grateful to the anonymous referee
for useful comments and to D. Takei, M. Tsujimoto, and S. Kitamoto
for stimulating discussion on V2491 Cyg.
This research has been supported in part by the Grant-in-Aid for
Scientific Research (20540227) 
of the Japan Society for the Promotion of Science.

\clearpage

\clearpage






\begin{thebibliography}{}


%


\bibitem[Baklanov et al. (2008)]{bak08}
Baklanov, A., Pavlenko, E., \& Berezina, E. 2008,
The Astronomer's Telegram, 1514




%


\bibitem[Bonifacio et al. (2000)]{bon00}
Bonifacio, P., Selvelli, P. L., \& Caffau, E. 2000, \aap,
356, L53










%






%

\bibitem[Dobrotka et al. (2006)]{dob06}
Dobrotka, A., Friedjung, M., Retter, A., Hric, L., \& Novak, R.
2006, \aap, 448, 1107














\bibitem[Goranskij et al. (2008)]{gor08}
Goranskij, P. V., Metlova, V. N., \& Burenkov, N. A. 2008,
The Astronomer's Telegram, 928, 1





\bibitem[Hachisu \& Kato (2001a)]{hac01ka}
Hachisu, I., \& Kato, M. 2001a, \apjl, 553, L161

\bibitem[Hachisu \& Kato (2001b)]{hac01kb}
Hachisu, I., \& Kato, M. 2001b, \apj, 558, 323


\bibitem[Hachisu \& Kato (2003b)]{hac03kb}
Hachisu, I., \& Kato, M. 2003b, \apj, 590, 445


\bibitem[Hachisu \& Kato (2004)]{hac04k}
Hachisu, I., \& Kato, M. 2004, \apj, 612, L57


\bibitem[Hachisu \& Kato (2006)]{hac06kb}
Hachisu, I., \& Kato, M. 2006, \apjs, 167, 59

\bibitem[Hachisu \& Kato (2007)]{hac07k}
Hachisu, I., \& Kato, M. 2007, \apj, 662, 552

\bibitem[Hachisu et al. (2008)Hachisu, Kato, \& Cassatella]{hac08kc}
Hachisu, I., Kato, M., \& Cassatella, A. 2008, \apj, 687, 1236


\bibitem[Hachisu et al. (2000)]{hkkm00}
Hachisu, I., Kato, M., Kato, T., \& Matsumoto, K. 2000, 
\apjl, 528, L97 



\bibitem[Hachisu et al. (2007)Hachisu, Kato, \& Luna]{hac07kl}
Hachisu, I., Kato, M., \& Luna, G. J. M. 2007, \apj, 659, L153

\bibitem[Hachisu et al. (1996)Hachisu, Kato, \& Nomoto]{hkn96}
Hachisu, I., Kato, M., \& Nomoto, K. 1996, \apj, 470, L97 

\bibitem[Hachisu et al. (1999a)Hachisu, Kato, \& Nomoto]{hkn99}
Hachisu, I., Kato, M., \& Nomoto, K. 1999a, \apj, 522, 487 

\bibitem[Hachisu et al. (1999b)]{hknu99}
Hachisu, I., Kato, M., Nomoto, K., \& Umeda, H. 1999b, \apj, 519, 314 

\bibitem[Hachisu et al. (2003)Hachisu, Kato, \& Schaefer]{hac03a}
Hachisu, I., Kato, M., \& Schaefer, B. E. 2003, \apj, 584, 1008
%







\bibitem[Ibarra \& Kuulkers (2008)]{iba08a}
Ibarra, A., \& Kuulkers, E. 2008, The Astronomer's Telegram, 1473, 1

\bibitem[Ibarra et al. (2008)]{iba08b}
Ibarra, A., et al. 2008, The Astronomer's Telegram, 1478, 1

\bibitem[Iglesias \& Rogers (1996)]{igl96}
Iglesias, C. A., \& Rogers, F. J. 1996, \apj, 464, 943



\bibitem[Kato (1983)]{kat83}
Kato, M. 1983, \pasj,  35, 507


\bibitem[Kato (1997)]{kat97}
Kato, M. 1997, \apjs, 113, 121

\bibitem[Kato (1999)]{kat99}
Kato, M. 1999, \pasj, 51, 525

\bibitem[Kato \& Hachisu (1994)]{kat94h}
Kato, M., \& Hachisu, I., 1994, \apj, 437, 802

\bibitem[Kato \& Hachisu (2007)]{kat07h}
Kato, M., \& Hachisu, I., 2007, \apj, 657, 1004





\bibitem[Kimeswenger et al. (2008)]{kim08}
Kimeswenger, S., Dalnodar, S., Knapp, A.. Schafer, J.,
Unterguggenberger, S., \& Weiss, S. 2008, \aap, 479, L51
























\bibitem[Munari et al. (2008)]{mun08}
Munari, U., et al. 2008, \aap, in press (arXiv:0810.2387)

\bibitem[Nakano et al. (2008)]{nak08}
Nakano, S., Beize, J., Jin, Z.-W., Gao, X., Yamaoka, H., Haseda, K.,
Guido, E., Sostero, G., Klingenberg, G., \& Kadota, K. 2008,
\iaucirc, 8934, 1

\bibitem[Nakano et al. (1999)]{nak99}
Nakano, S., Tago, A., \& Nakamura, A. 1999, \iaucirc, 7223, 1









\bibitem[Page et al. (2009)]{pag09}
Page, K. L., et al. 2009, in preparation




\bibitem[Payne-Gaposchkin (1957)]{pay57}
Payne-Gaposchkin, C. 1957, The Galactic Novae (Amsterdam:
North-Holland)


































\bibitem[Takei et al. (2009)]{tak09}
Takei, D., Tsujimoto, M., Kitamoto, S., Ness, J.-U., Drake, J. J.,
\& Takahashi, H. 2009, in preparation


\bibitem[Tomov et al. (2008)]{tom08}
Tomov, T., Mikolajewski, M., Brozek, T., Ragan, E., Swierczynski, E.,
Wychudzki, P., \& Galan, C. 2008, The Astronomer's Telegram, 1485



\bibitem[Venturini et al. (2004)]{ven04}
Venturini, C. C., Rudy, R. J., Lynch, D. K., Mazuk, S., \& Puetter, R. C.
2004, \aj, 128, 405


\bibitem[Warner (1995)]{war95}
Warner, B. 1995, Cataclysmic variable stars, Cambridge, 
Cambridge University Press

%



%
%
\end{thebibliography}
\end{document}